\documentstyle[epsfig]{l-aa}
\def\jref#1 #2 #3 #4 {{\par\noindent \hangindent=2em \hangafter=1
      \advance \rightskip by 0em #1, {\it#2}, {\bf#3}, #4.\par}}
\def\rref#1{{\par\noindent \hangindent=2em \hangafter=1
      \advance \rightskip by 0em #1.\par}}

\def\wisk#1{\ifmmode{#1}\else{$#1$}\fi}

\def\deg{\ifmmode^{\circ}\else$^{\circ}$\fi} 
\def\min{\ifmmode^{\prime}\;\else$^{\prime}\;$\fi}
\def\sec{\ifmmode^{\prime\prime}\;\else$^{\prime\prime}\;$\fi}

\def\dn{\ifmmode{\Delta\nu{_d} }\else{$\Delta\nu_{d}$ }\fi}
\def\dt{\ifmmode{\Delta t{_d} }\else{$\Delta t_{d}$ }\fi}
\def\lsim{\,\lower2truept\hbox{${< \atop\hbox{\raise4truept\hbox{$\sim$}}}$}\,}
\def\gsim{\,\lower2truept\hbox{${> \atop\hbox{\raise4truept\hbox{$\sim$}}}$}\,}

\onecolumn
\title{On the detection of neutrino oscillations with Planck surveyor}
\author{{L. Popa}\inst{1,2} \and {C. Burigana}\inst{2}
\and {F. Finelli}\inst{2,3}\footnote{Present Address: Physics Dept.,
Purdue University,
West Lafayette,
IN 47907-1396, USA}
\and {N. Mandolesi}\inst{2} }
\begin{document}
\offprints{lpopa@venus.nipne.ro or popa@tesre.bo.cnr.it}
\date{Received....2000 / Accepted....2000}
\thesaurus{12(12.03.1; 12.04.1; 12.12.1) 02(02.05.1)}
\institute{
{Institute of Space Sciences, Bucharest-Magurele, R-76900, Romania}
\and
{Istituto TeSRE, Consiglio Nazionale delle Ricerche, Via Gobetti 101,
I-40129 Bologna, Italy}
\and
{University of Bologna and INFN, Via Irnerio 46, I-40126 Bologna, Italy}}

\maketitle
\markboth{L. Popa et al.: On the detection of neutrino oscillations with
Planck surveyor}
{L. Popa et al.: On the detection of neutrino oscillations with
Planck surveyor}

\begin{abstract}
The imprint of  neutrino  oscillations
on the Cosmic Microwave Background (CMB)
anisotropy and polarization power spectra is evaluated 
in a $\Lambda$CHDM model
with two active neutrino flavors, consistent with the structure formation
models and the atmospheric neutrino oscillations data.\\
By using the Fisher information matrix method we find that
the neutrino oscillations
could be detected by {\sc Planck} surveyor if the present value 
of the lepton asymmetry 
$L_{\nu} \geq 4.5 \times 10^{-3}$,
the difference of the neutrino
squared masses $\Delta m^2 \geq  9.7 \times 10^{-3}$eV$^2$
and the vacuum mixing angle
$\sin^2 2\theta_0 \geq 0.13$, showing the existence of a significant
overlap between the region of the
oscillation parameter space that can be measured by {\sc Planck}
surveyor and that implied by the atmospheric neutrino oscillations data.\\


\end{abstract}

\section{Introduction}

The atmospheric neutrino results
from   Super-Kamiokande (Fukuda et al. 1998)
and MACRO (Ambrosio et al. 1998) experiments
indicate that neutrinos oscillate.
Those
data are consistent with $\nu_{\mu} \leftrightarrow \nu_{\tau}$
oscillations, but do not exclude $\nu_{\mu}$ oscillations to
a sterile neutrino $\nu_{s}$
(see e.g. Foot et al. 1996; Foot \& Volkas 1997,1999).
The small value of the difference of the squared masses
($5 \times 10^{-4}{\rm eV}^2  \leq \Delta m^{2}
\leq 6 \times 10^{-3}{\rm eV}^2$)
and the strong mixing angle ($ \sin^{2} 2\theta \geq 0.82$) suggest that
these neutrinos are nearly equal in mass
as predicted by many models of particle physics beyond the
standard model (see e.g. Primack \& Gross 1998 and the references
therein).
Also, the LSND experiment (Athanassopoulos et al. 1998) support
$\nu_{\mu}\leftrightarrow \nu_{e}$ oscillations with
$\Delta m^{2} \leq 0.2$eV$^{2}$ and
other different types of solar neutrino experiments
(Bahcall et al. 1998) suggest that $\nu_e$ could oscillate to a sterile
neutrino $\nu_e \leftrightarrow \nu_s$ with $\Delta m^{2} \simeq 10^{-5}$eV$^2$.

The direct implication of neutrino oscillations is the existence of
non-zero neutrino masses in the eV range, and consequently
a not negligible hot dark matter contribution
to  the total mass density of the universe
(i.e. a density parameter $\Omega_{\nu}~\neq 0$).

In view of the uncertainty in the direct neutrino experiments,
a possible significant contribution can be obtained from the study of
the cosmological implications of neutrino oscillations.
In particular, the Cosmic Microwave Background (CMB) anisotropy pattern
reflects the conditions in the universe at the time
of the last scattering between photons and electrons,
occurring at a redshift $z\approx 1000$ for standard recombination models.
This means that all the physical processes
occurring before this epoch could have left imprints on the CMB
angular power spectra.

The standard Cold Dark Matter (CDM) model
normalized to  COBE/DMR data
(Smoot et al.~1992, Wright et al.~1994,
G\'orski et al.~1994, Bennet et al.~1996a)
predicts both the  amplitude and the shape of the CMB
power spectrum at small scales
inconsistent with the observations
of the Large Scale Structure (LLS) of the universe as derived by galaxy
surveys (e.g. Scott \& White 1994; White et al.~1995;
Primack et al.~1995).  
Recent works (Primack 1998; Gawiser \& Silk 1998) show
that the Cold + Hot Dark Matter (CHDM) model
is the single model with $\Omega_{m}=1$
whose predictions agree with both
LSS observations and current CMB anisotropies data.
The C$\nu^2$DM model
(Primack et al. 1995) with two $2.4$eV neutrinos
contributing with $20\%$ to the total density
of the universe and  with another $80\%$ contribution from cold dark matter
and a small baryon fraction,
agrees remarkably with all available
observations  only if the Hubble parameter is $H_0= 50$ km s$^{-1}$Mpc$^{-1}$
($h_0=H_0/100$ km s$^{-1}$Mpc$^{-1}=0.5$).

However, increasing evidence for larger value of the Hubble parameter,
$h_0 \simeq 0.6-0.8$, (see e.g. Fukugita, Liu \& Sugiyama 1999
and the references therein) are
inconsistent with standard CHDM models. 
Such high values of the Hubble parameter
make an $\Omega_m=1$ universe suspiciously young unless the cosmological 
constant is non-zero.
A way to improve the agreement of the standard
CHDM models with eV neutrino mass with high values of Hubble parameter is
to consider the relic neutrino degeneracy (the degeneracy parameter is 
defined as: 
$\xi_{\nu}=\mu_{\nu}/T_{\nu}$, where $\mu_{\nu}$ is the neutrino chemical 
potential and $T_{\nu}$ is the neutrino temperature),    
that enhances the contribution of neutrinos to the total energy density of the universe
(Larsen \& Madsen 1995).
The cosmological implications of the neutrino degeneracy has been often 
considered in the literature: it can change the neutrino decoupling temperature
(Freese et al. 1983; Kang \& Steigman 1992), the abundances of light elements
at the big bang nucleosynthesis (BBN)
(Steigman et al. 1977; Kang \& Steigman 1992), the CMB anisotropies and the
matter power spectrum (Kinney \& Riotto 1999). For the CHDM models
with degenerated neutrinos, the increase of the neutrino chemical 
potential increases the effective number of neutrino species
in the relativistic era (Larsen \& Madsen 1995, Hannestad 2000), 
and can bring  the power spectral shape parameter in the 
range required by  observations
[for CDM models with low baryonic content 
the power spectral shape parameter (which basically measures the horizon
scale
at matter-radiation equality)
is defined  as (Dodelson et al. 1994):
$\Gamma \approx \Omega_m h_0 (g_*/3.36)^{-1/2}$, where $g_*$ counts the 
relativistic degrees of freedom and $g_*=3.36$ corresponds to the standard 
model with photons and three massless neutrino species; observations 
require $0.22< \Gamma < 0.29$].
The free-streaming properties of degenerated neutrinos also differ: 
when the neutrino chemical potential is increased,  
the neutrino number density is higher, 
although the neutrino energy density 
and the mean momentum change a little. 
The effect is the increase of  the 
radiation energy density and the
delay of the matter-radiation equality. 
It is shown (Larsen \& Madsen 1995) that 
an $\Omega_m=1$ CHDM model with one $2.4$eV neutrino and $h_0=0.7$ 
is in good agreement with the observed LSS power spectrum of 
density fluctuations  if the neutrino
degeneracy parameter is $\xi_{\nu} \approx 2.8$. 
Constraints on neutrino degeneracy coming from BBN (Kang \& Steigman 1992) 
indicate $-0.06 \leq \xi_{\nu_e} \leq 1.1$ 
and  $|\xi_{\nu_{\mu,\tau} }| \leq 6.9$.
Recent works (Pal \& Kar 1999; Lesgourgues \& Pastor 1999; Kinney \& Riotto 
1999, Lesgourgues et al. 2000) derive  
bounds on neutrino degeneracy parameter
in agreement with BBN predictions, 
by combining the current observations of the CMB anisotropies and
LSS data.  The likelihood analysis of the CMB anisotropy 
data obtained by the Boomerang experiment
indicates bounds on massless neutrino degeneracy parameter
(Hannestad 2000) of
$|\xi_{\nu_{e,\mu,\tau}}| \leq 3.7$ if only one massless 
neutrino species is degenerated and  $|\xi_{\nu}| \leq 2.4$ if the 
asymmetry is equally shared among three massless species.\\
On the other hand, evidences has been accumulated that we live  
in a low matter density universe (see e.g. Fukugita, Liu \& Sugiyama 1999
and the references therein).   
Indications like Hubble diagram of Type 1a 
supernovae (Riess et al. 1998, Perlmutter et al. 1998) and 
the acoustic peak distribution in the CMB anisotropy power spectra
(Hancock et al. 1998, Efstathiou et al. 1999) point to a
universe dominated by vacuum energy (cosmological constant
$\Lambda$) that keeps the universe close to flat.
The combined analysis of the latest CMB anisotropy data 
and Type 1a supernovae data (Efstathiou et al. 1999)
indicates $\Omega_m=0.25^{+0.18}_{-0.12}$ and 
$\Omega_{\Lambda}=0.63^{+0.17}_{-0.23}$ ($95$\% confidence errors)
for matter and vacuum  energy densities respectively.
These values are close to those favoured by other arguments 
(see e.g. Efstathiou et al. 1999 and the references therein) like the ages of
globular clusters, observations of 
large scale structure, baryon abundance in clusters.\\
Adding a Hot Dark Matter component to the $\Lambda$CDM model
($\Lambda$CHDM) leads to a worse fit to LSS and CMB data, 
resulting in a limit on the total neutrino mass (Gawiser 2000)
of $m_{\nu} \leq 2$eV
for a primordial scale invariant power spectrum 
and $m_{\nu} \leq 4$eV for a primordial scale free power spectrum. 
A stronger upper limit is obtained 
(Fukugita, Liu \& Sugiyama 1999) from 
the matching condition of the LSS 
power spectrum normalization $\sigma_8$ (defined as the {\it rms} amplitude
of the galaxy power spectrum in a sphere of radius 8$h_0^{-1}$Mpc)   
at the COBE
scale and at the cluster scale. 
For the case of a $\Lambda$CHDM model having 
$\Omega_m=0.3$, $\Omega_{\Lambda}=0.7$, and a
primordial scale invariant power spectrum, it is found an upper limit 
of the total non-degenerated neutrino mass of $m_{\nu} < 0.6$  if the Hubble
constant $H_0 < 80$ Km s$^{-1}$ Mpc$^{-1}$.
 
In this paper we study the signature of relic degenerated neutrino
oscillations on the CMB anisotropy and polarization power spectra
and address its detectability with the future CMB anisotropy
space missions, MAP (Microwave Anisotropy Probe)
(see Bennet et al. 1996b) and {\sc Planck} (Mandolesi et al. 1998,
Puget et al. 1998).
We consider the impact of neutrino oscillations 
in the epoch after nucleosynthesis.
Neutrino oscillations are mediated by weak
interactions: this fact implies that the order of magnitude of the
interaction rate is less than the Hubble
expansion rate after nucleosynthesis. In order to get an appreciable
effect we consider a large neutrino asymmetry as left by processes
occurred before and during nucleosynthesis. An example of such large
asymmetries is given by considering a relic neutrino degeneracy (Larsen
and Madsen 1995), which
has been recently subject of renewed interest (Kinney and Riotto 1999;
Lesgourges and Pastor 1999; Pal and Kar 1999). 
We analyze the oscillations of relic degenerated neutrinos under the simple
assumption of oscillations occurring between two active neutrino flavors
\begin{eqnarray}
\nu_{\mu} \leftrightarrow  \nu_{\tau}  \hspace{0.5cm}
\bar{\nu}_{\mu}  \leftrightarrow  \bar{\nu}_{\tau}  \nonumber
\end{eqnarray}
which have the following mass hierarchy:
\begin{eqnarray}
m_{\nu_{\tau}} = m_{\bar{\nu}_{\tau}} > \;
m_{\nu_{\mu}} = m_{\bar{\nu}_{\mu}} \nonumber
\end{eqnarray}
We assume the third neutrino $\nu_{e}$ as massless and non-degenerate.
We consider a total neutrino mass contribution of 
$m_{\nu}=m_{\nu_{\mu}}+ m_{\nu_{\tau}}= 0.6$ eV and a total neutrino
degeneracy parameter
$\xi_{\nu}=|\xi_{\nu_{\mu}}+\xi_{\nu_{\tau}}|\leq 5$ accordingly to the
present observational data summarized above.
We  assume a scale invariant primordial power spectrum, the presence of the 
scalar modes with spectral index $n_s=1$ 
and ignore the contribution 
of the tensorial modes and the reionisation effects. 
This model is consistent with the large scale structure and CMB anisotropy 
data, allowing in the same time a pattern of  neutrino masses 
consistent with the results from
atmospheric neutrino oscillations experiments.\\
In Section 2 we draw the basic formalism of the neutrino oscillation model
necessary for understanding its impact on CMB angular power spectra.
The features induced by the degenerated
neutrino oscillations on the CMB anisotropy and polarization
power spectra are presented in Section 3.
In Section 4 we discuss the MAP and {\sc Planck} capability
of detecting neutrino oscillations. Finally, we summarize
our main conclusions in Section 5.\\
Throughout the paper we employ the system of units
in which $\hbar=c=k_B=1$.

\section{Oscillations of degenerated neutrinos}

The neutrino oscillations take place due to the fact that
neutrino mass eigenstate components propagate differently because
they have different energies, momenta and masses.
In the standard assumption of the mixing of massive neutrinos,
the neutrino flavor eigenstates are described by a 
superposition of the mass eigenstate components (Particle Data Group 1998).
For  oscillations occurring in vacuum between two neutrino flavors
$\nu_{\mu}$ and $\nu_{\tau}$ the mixing can be written as:
\begin{eqnarray}
| \nu_{\mu} \rangle & = &\;\;\cos \theta_0 | \nu_{2} \rangle + \sin
\theta_0 | \nu_{3} \rangle \\
| \nu_{\tau} \rangle &= &-\sin \theta_0 | \nu_{2} \rangle + \cos \theta_0
| \nu_{3} \rangle,
\nonumber
\end{eqnarray}
where  $\nu_2$ and $\nu_3$ are the mass eigenstate components and
$\theta_0$ is the vacuum mixing angle. The mixing of antineutrinos
can be obtained from equation (1) by performing the transformations
$\nu_{\mu} \leftrightarrow  {\bar \nu}_{\mu}$ and
$\nu_{\tau} \leftrightarrow {\bar \nu}_{\tau}$.
It is usual to consider that each neutrino/antineutrino
of a definite flavor is dominantly one mass eigenstate
(Particle Data Group 1998).
In this circumstance we refer to the dominant mass eigenstate
component of $\nu_{\mu}$/${\bar\nu}_{\mu}$
as $\nu_2$, that of $\nu_{\tau}$/${\bar \nu}_{\tau}$ as $\nu_3$
and to their difference of the squared masses
as $\Delta m^2=m^2_3-m^2_2$.

The mixing of the mass eigenstate components are modified in the presence of
the asymmetric background. The background mixing angle and the vacuum mixing
angle are related through (see e.g. Foot et al. 1996):
\begin{equation}
\sin^2 2\theta_m=\frac{\sin^2 2\theta_0}{1-2z \cos 2\theta_0+z^2}.
\end{equation}
Here $z=2<p> <V_{\nu_i}>/\Delta m^2$ ($i=\mu,\tau$),
$<p>$ is the averaged neutrino momentum and
$<V_{\nu_i}>$ is the neutrino effective potential
due to the interaction with the asymmetric background
(Enqvist et al. 1992; Foot et al. 1996):
\begin{equation}
<V_{\nu_i}>=\sqrt 2 G_F n_{\gamma} \left( L^{i}-A_{\nu_i}\frac{pT_{\nu}}{4M^2_Z}
[n_{\nu_i}+n_{{\bar \nu_i}}]  \right),
\end{equation}
where $A_{\nu_i}$ is a numerical factor
($A_{{\mu},{\tau}}=12.61$),
$T_{\nu}$ is the neutrino temperature,
$G_F$ is the Fermi constant ($G_F=1.17 \times 10^{-11}MeV^{-2}$)
$M_Z$ is the mass of $Z$ boson,
($M_Z=91.187$GeV), $n_{\gamma}$ is the photon number density
($n_{\gamma} \simeq T^3/4.1$),
$n_{\nu_i}$ and $n_{\bar{\nu_i}}$ are the neutrino and antineutrino
number densities.\\
The functions $L^{i}$ are defined as (Foot et al. 1996;
Foot \& Volkas 1997):
\begin{eqnarray}
L^{i}=L_{\nu_i}+ L_{\nu_{\mu}}+ L_{\nu_{\tau}}+L_{\nu_{e}} + L_{e}+\eta
\hspace{1.3cm} (i=\mu,\tau).
\end{eqnarray}
Here $\eta$ is approximately equal to the baryon to photon ratio,
and $L^{'}$s are the lepton asymmetries defined as:
\begin{equation}
L_{\alpha}=\frac{ n_{\alpha} - n_{{\bar \alpha}} } {n_{\gamma}},
\end{equation}
where $n^{'}$s are the number densities.\\
For the purpose of this work we consider large neutrino asymmetries ($L_{\nu_i} \geq 10^{-9}$).
Also, we consider that $\nu_{e}$ is non-degenerated and consequently
$L_{\nu_e}=0$.
In this case the functions $L^i$ may be written as:
\begin{equation}
L^{i} \simeq L_{\nu_i}+L_{\nu_{\mu}}+L_{\nu_{\tau}}.
\end{equation}
The effective potentials of antineutrinos can be obtained by replacing
$L^i$ with $-L^i$ in  equation (3).

The neutrino flavor eigenstates oscillate via weak
interaction processes. The average oscillation probability
$<{\cal{P}}>_{\nu_i}$
of a flavor eigenstate $|\nu_i \rangle$ ($i=\mu,\tau$) 
is defined as 
(see e.g.: Raffelt 1996, Foot et al.1996, Enqvist et al. 1992):
\begin{eqnarray}
< {\cal{P}} >_{\nu_i}=
\sin^2 2 \theta_{m} \left < \sin^2 \frac {L_{int}} {L_{osc}} \right>
\Gamma_{coll}(\nu_{i} \rightarrow \nu_j)
\hspace{1cm} (i=\mu,\tau \;\; j=\mu,\tau\;\; i \not= j),
\end{eqnarray}
where $\Gamma_{coll}(\nu_i \rightarrow \nu_j)$
is the neutrino averaged elastic collision rate, $L_{int}$ is the
mean distance between interactions (Foot et al. 1996), and the neutrino
oscillation length 
$L_{osc}$ is given by (see e.g. Raffelt 1996):
\begin{equation}
L_{osc}  =\frac{4\pi E_{\nu_i}}{\Delta m^2}
\simeq 248 {\rm cm} \frac{E_{\nu_i}}{\rm MeV} 
\frac {{\rm eV}^2}{\Delta m^2}  \, ,
\end{equation}
where $E_{\nu_i}$ is the neutrino energy. We will consider the
$L_{int}>> L_{osc}$ and then  
$\left < \sin^2 (L_{int}/ L_{osc}) \right> \rightarrow 1/2$ 
(Foot et al. 1996). 
The averaged collision rate is given by (Enqvist et al. 1992):
\begin{equation}
\Gamma(\nu_i  \rightarrow \nu_j) =
                 0.77G_F^2 T_{\nu}^5 \frac{n_{\nu_i}}{n_0},
\end{equation}
where $n_0$
is the number density of massless non-degenerated neutrino species.

Equation (7) together with equations (2), (8), (9) gives
the averaged oscillation probability $<{\cal{P}}>_{\nu_i}$ used in the
next section to evolve in time the neutrino distribution functions. 
The oscillation probabilities of antineutrinos can be obtained by performing the
transformations $\nu_i \leftrightarrow {\bar \nu_i}$
and  $\nu_j \leftrightarrow {\bar \nu_j}$
in equations $(7) - (9)$.  

\section{CMB angular power spectra in the presence of neutrino oscillations}

The phase space distribution functions of neutrinos and antineutrinos
$f_{\nu_i}$ ($i=\nu_{\mu},{\bar \nu}_{\mu},
\nu_{\tau},{\bar \nu}_{\tau}$)
are evolved independently, taking into account the  redistribution of
momenta due to oscillations and elastic collisions.\\
We describe the time evolution of the phase space distribution functions
in the expanding universe  by the set of equations:
\begin{eqnarray}
\frac{\partial f_{\nu_i}(q)}{\partial t} & = &
-< {\cal P} >_{\nu_i}  f_{\nu_i}(q)+ < {\cal P } >_{\nu_j}  f_{\nu_j}(q),
\hspace{2.5cm} (i=\mu,\tau; \;\; j=\mu,\tau; \;\; i \not= j) \nonumber \\
\frac{\partial f_{{\bar \nu}_i}(q)}{\partial t} & = &
-< {\cal P} >_{ {\bar \nu}_i}  f_{{\bar\nu}_i}(q)+ 
< {\cal P } >_{{\bar \nu}_j}  f_{{\bar\nu}_j}(q), 
\end{eqnarray}
where $q=ap$ is the comoving momentum (Ma \& Bertschinger 1995),
$p$ is the magnitude of the momentum
3-vector, $a$ is the scale factor ($a=1$ today), $dt=da/(aH)$ and $H$ is
the expansion rate:
\begin{equation}
H=\sqrt{ \frac{8 \pi G \rho_{tot}(a)}{3}} \, ,
\end{equation} 
where: $G$ is the gravitational constant and 
$\rho_{tot}(a)$ is the total energy density.\\ 
The equations (10) assume that
no other physical processes take place
apart from oscillations and elastic collisions,
leading to the conservation of the total neutrino/antineutrino
number densities $(n_{\nu}$/$n_{{\bar \nu}})$ 
and degeneracy parameters $(\xi_{\nu}$/$\xi_{{\bar\nu}})$:
\begin{eqnarray}
n_{\nu}=n_{\nu_i}+n_{\nu_j},\;\;\;\;
n_{{\bar \nu}}&=&n_{{\bar\nu}_i}+n_{\bar{\nu}_j}, 
\hspace{5.5cm}     (i=\mu,j=\tau)\nonumber\\ 
\xi_{\nu}=\xi_{\nu_i}+\xi_{\nu_j},\;\;\;\;
\xi_{{\bar\nu}}&=&\xi_{{\bar \nu}_i}+\xi_{{\bar\nu}_j},
\end{eqnarray}
where $\xi_{\alpha}=\mu_{\nu_{\alpha}}/T_{\nu}$ 
is the degeneracy parameter of the flavor $\alpha$ $(\alpha=\mu,\tau)$
with the chemical potential $\mu_{\alpha}$,
${T_{\nu}}$ is the neutrino/antineutrino temperature and  
$\vert {\xi_{{\bar \nu}_{\alpha}}}|={\vert \xi_{\nu_{\alpha}}}|$.

We have modified the CMBFAST code 
(Seljak \& Zaldarriaga 1996) 
to take into account the evolution of 
the phase space distribution
functions of neutrinos and antineutrinos at each value of the scale factor,
by solving the set of equations (10).
We start at the scale factor $a \simeq 10^{-9}$,
when  neutrinos and antineutrinos with the mass in eV range behave like
relativistic particles with Fermi-Dirac phase space distributions:
\begin{eqnarray}
f_{\nu_i}(q)=\frac{1} {e^{E_{\nu_i}/T_\nu -\xi_{\nu_i} }+1 },  \;\;\;
f_{\bar{\nu}_i}(q)=\frac{1} {e^{E_{\nu_i}/T_\nu +\xi_{\nu_i} }+1 },
\hspace{5cm} (i=\mu,\tau)
\end{eqnarray}
where $\xi_i$ is  the degeneracy parameter, $E_{\nu_i}=\sqrt{q^2+
a^2 m^{2}_{\nu_i}}$  the neutrino/antineutrino
energy, $m_{\nu _i}$  the dominant mass component of the flavor $i$,
and $q$ the comoving momentum.
The system of neutrinos and antineutrinos
drops out from thermal equilibrium with $e^{\pm}$,
photons, the small fraction of baryons and other massive species
(e.g. $\mu^+ \mu^-$ pairs) when the ratio of the averaged weak
interaction rate
to the expansion rate falls below unity.
The decoupling temperature of degenerated neutrinos increases
exponentially
with the degeneracy parameter. For  $\nu_{\mu}$ and $\nu_{\tau}$  
the decoupling temperature
is given by (Freese et al. 1983, Kang \& Steigmann 1992)
$T_{D} \simeq 10$ MeV $\xi^{-2/3}e^{(\xi/3)}$.\\
The present values of neutrino/antineutrino temperature $T_{\nu_0}$
and photons temperature $T_0=2.725\pm 0.002$K (Mather et al. 1999)
are related through  $T_{\nu_0}/T_{0}\simeq (3.9/g_{*D})^{1/3}$
where $g_{*D}$ is the number
of degrees of freedom in equilibrium at 
neutrino/antineutrino decoupling. For a degeneracy parameter 
$\xi_{\nu} \leq 15$, $T_{D}\simeq$ few MeV and
the total number of relativistic degrees of freedom 
is $g_{*}\simeq 43/4$ (Freese et al. 1983).
After the neutrino decoupling, the $e^+e^-$ pairs annihilate,
heating the photons but not the neutrinos.
From the entropy conservation one obtains that
$T_{\nu_0}\simeq (4/11)^{1/3}T_0$
and
the total degeneracy parameter \
$\xi_{\nu}=\xi_{\nu_{\mu}}+\xi_{\nu_{\tau}}$ 
is considered constant.

We compute  the mean energy density and pressure of each neutrino
flavor 
as:
\begin{eqnarray}
\rho_{\nu_{i}}+\rho_{\bar{\nu}_{i}}&=&
\frac{T_{\nu}^4 }{2 \pi^2} \int^{\infty}_0 dq \; q^2 E_{\nu_i}
(f_{\nu_i}(q)+f_{ \bar{\nu}_i}(q)), \hspace{3cm} (i=\mu,\tau)\nonumber \\
P_{\nu_{i}}+P_{\bar{\nu}_{i}}&=&
\frac{T_{\nu}^4 }{6 \pi^2} \int^{\infty}_0 dq \; \frac{q^2}
{E_{\nu_i}} (f_{\nu_i}(q)+f_{ \bar{\nu}_i}(q)),
\end{eqnarray}
the  neutrino and antineutrino number densities: 
\begin{eqnarray}
n_{\nu_i} =  \frac{T_{\nu}^4 }{2 \pi^2}
\int^{\infty}_0 dq \; q^2 f_{\nu_i}(q), \;\;\;\;
n_{{\bar \nu}_i} &=& \frac{T_{\nu}^4 }{2 \pi^2}
\int^{\infty}_0 dq \; q^2 f_{{\bar \nu}_i}(q), \hspace{2cm} (i=\mu,\tau)
\end{eqnarray}
and the present lepton asymmetry in the form of neutrinos (Kang \& Steigmann 1992):
\begin{equation}
L_{\nu}=\sum_{i=\mu,\tau}
\frac{1}{12 \zeta(3)} \frac {T_{\nu_0}}{T_0}[\xi_{\nu_i}+\pi^2 \xi_{\nu_i}].
\end{equation}
We  also compute at the beginning of the code
the present value of neutrino parameter density:
\begin{equation}
\Omega_{\nu}=
\sum_{i=\mu,\tau} \frac{( \rho_{\nu_{i}}+ \rho_{\bar{\nu}_{i}})_{a=1}}{\rho_c},
\end{equation}
where $\rho_c=1.054 \times 10^4 h^2_0$ eV cm$^{-3}$
is the critical  density, and
define the cold dark matter density parameter 
$\Omega_c=1-\Omega_b-\Omega_{\nu}-\Omega_{\Lambda}$.\\
Following the same procedure used by the CMBFAST code,
we compute in synchronous gauge the perturbations of the
energy density, pressure, energy flux and shear stress 
(equations $(52) - (56)$ from  Ma \& Bertschinger 1995)  
for neutrinos and antineutrinos of each flavor.\\
We performed the computations of the CMB anisotropy and polarization
power spectra starting with  $N_{q_{max}}=15$ equispaced 
sampling points of the phase 
space distribution functions and truncating the Boltzmann hierarchies
for massive neutrinos (see equation (54) from Ma \& Bertschinger 1995)  
at $l_{max}=15$ for every $q$ value.
We then gradually increase these numbers
until $N_{q_{max}}=100$ and $l_{max}=50$ 
for testing the numerical precision and 
obtaining
a relative accuracy better than $10^{-3}$ for the
models with oscillations and without oscillations.
\begin{figure*}
\picplace{10cm}
\caption{The dependence of CMB anisotropy (left panels) and polarization
(right panels) power spectra  on $\Delta m^2$ ( panels a1) and a2) )
$\sin^2 2 \theta_0$ ( panels b1) and b2) ) 
and $\Delta \xi_{\nu}$ ( panels c1) and c2) ) for a $\Lambda$CHDM model
having: $\Omega_b=0.023$, $\Omega_{\Lambda}=0.7$,
$\Omega_c=1-\Omega_b-\Omega_{\nu}-\Omega_{\Lambda}$,
$h_0=0.65$, $m_{\nu}=m_{\nu_{\mu}}+m_{\nu_{\tau}}=0.6$eV,
$\xi_{\nu}=\xi_{\nu_{\mu}}+\xi_{\nu_{\tau}}=5$.
As a consistency test we present in 
panel a1) the CMB anisotropy
power spectra obtained for degenerate neutrinos with: $m_{\nu}=0.07$eV,
$\xi_{\nu}=3$ (dashed line),  $m_{\nu}=0.07$eV,  $\xi_{\nu}=0$
(dash-dotted line) and  $ h_{0}=0.65$, $\Omega_b=0.05$, $\Omega_{\Lambda}=0.7$,
$\Omega_c=1-\Omega_b-\Omega_{\nu}-\Omega_{\Lambda}$.
In panels b1) and b2) we present (dashed lines)
the CMB angular power spectra obtained for:
$\Omega_{\nu}\approx 0.015$, $\xi_{\nu}=0$, $\Omega_{b}=0.023$,
$\Omega_{\Lambda}=0.7$,  $\Omega_c=0.262$, $h_0=0.65$, one massless
and two massive neutrino species
(see also the text). We choose such values of
neutrino oscillation parameters in order to 
clearly see the neutrino oscillation effects on CMB power spectrum. }
\end{figure*}

Figure 1 presents the dependence of the CMB anisotropy
and polarization power spectra on the neutrino difference of the 
squared masses  $\Delta m^2$, the vacuum mixing angle 
$\sin^2 2 \theta_0$ and  the lepton asymmetry $L_{\nu}$,  
obtained in a $\Lambda$CHDM model having:
$\Omega_b=0.023$, $\Omega_{\Lambda}=0.7$,
$\Omega_c=1-\Omega_b-\Omega_{\nu}-\Omega_{\Lambda}$,
$h_0=0.65$.
For all cases  we assume a total neutrino mass
$m_{\nu}=m_{\nu_{\mu}}+m_{\nu_{\tau}}=0.6$eV and a total neutrino 
degeneracy parameter $\xi_{\nu}=\xi_{\nu_{\mu}}+\xi_{\nu_{\tau}}=5$.
We also  assume a  primordial scale-invariant
power spectrum, the presence of the scalar modes with
spectral index $n_s=1$ and a standard ionization history of the 
universe. All the power spectra are normalized to COBE/DMR four-Year
data (Bunn \& White 1997).
Panels a1) and a2) present the CMB anisotropy and polarization power spectra
for $\sin^2 2\theta_0=0.8$, $\Delta \xi_{\nu}=\xi_{\nu_{\tau}}-\xi_{\nu_{\mu}}=4$~
 ($L_{\nu}=3.54$) and some values of $\Delta m^2$.
Panels b1) and b2) show the same power spectra
for $\Delta m^2=0.24$eV$^2$, $\Delta \xi_{\nu}=4$,
and some values of $\sin^2 2 \theta_0$.
Panels c1) and c2) present the CMB anisotropy and polarization power 
spectra for $\Delta m^2=0.24$eV$^2$,  
$\sin^2 2\theta_0=0.45$ and $\Delta \xi_{\nu}$/$L_{\nu} =$ 0/2.03,
1/2.13, 3/2.88, 4/3.54. For all cases we indicate the 
corresponding values of $\Omega_{\nu}$
(equation 17). 
Without oscillations our power spectra are
in agreement with the CMB anisotropy power spectra 
presented  by Lesgourgues \& Pastor 1999.
Also, in panels b1) and b2) we present (dashed lines) 
the CMB angular power spectra obtained for: 
$\Omega_{\nu}\approx 0.015$, $\xi_{\nu}=0$, $\Omega_{b}=0.023$, 
$\Omega_{\Lambda}=0.7$,  $\Omega_c=0.262$, $h_0=0.65$, one massless 
and two massive neutrino species. 
We check that our results match exactly 
those obtained by using the unmodified version of the CMBFAST code.\\     
\begin{figure*}
\picplace{10cm}
\caption{The evolution with the scale factor of the energy
densities, relative to the critical density, of the various components,
for the models presented in Figure 1.
Panel a): $\sin^2 2\theta_0=0.8$, $\Delta \xi_{\nu}=4$
($L_{\nu}=3.54$) and $\Delta m^2$: 0.24eV$^2$ (continuous lines),
0.2eV$^2$ (long-dashed lines),  0.12eV$^2$ (dashed lines), 0 (dott-dashed lines).
Panel b): $\Delta m^2=0.24$eV$^2$, $\Delta \xi_{\nu}=4$
($L_{\nu}=3.54$) and $\sin^2 2\theta_0$: 0, 0.23, 0.45, 1 (see also the text).
Panel c): $\Delta m^2=0.24$, $\sin^2 2\theta_0=0.45$ 
and $\Delta \xi_{\nu}$/$L_{\nu}$: 4/3.54 (continuous lines), 3/2.88 
(long-dashed lines),
1/2.13 (dashed lines), 0/2.03 (dott-dashed lines). 
Panels a1), b1) and c1) show 
the separate contributions from the two types of massive neutrinos
(together with antineutrinos) for the same cases of the above panels.
For all values of the scale factor, $\sum_i \Omega_i=1$.}
\end{figure*} 
The power spectra presented in Figure 1 show two distinct features:
a vertical shift of the $C_l$ at large $l$ (present in all panels)
with the increasing of $\Delta m^2$, $\sin^2 2\theta_0$ and 
$\Delta \xi_{\nu}$ values,
that results from the modification of neutrino free streaming scales, and
an horizontal shift of the Doppler peaks to lower $l$ when 
$\Delta m^2$ and $\Delta \xi_{\nu}$
are increased, that results from the increase of the sound horizon at recombination.
The horizontal shift of $C_l$ is not present in panels b1) and b2)
because the variation of $\sin^2 2 \theta_0$, with $\Delta m^2$ and
$\Delta \xi_{\nu}$
kept fixed, does not change $\Omega_\nu$ and consequently the Hubble
expansion rate is not changed. 
(see equations 14 and 15).

For a better understanding of these features,  
Figure 2 shows the evolution with the scale factor of the energy
densities, relative to the critical density, of the various components
($\Omega_{cdm}$ for CDM, $\Omega_{bar}$ for baryons, $\Omega_{rad}$ for radiation
in the form of one massless neutrino and photons, $\Omega_{\nu_{\mu}}$
and $\Omega_{\nu_{\tau}}$
for the massive neutrinos and antineutrinos, $\Omega_{vac}$
for vacuum),
for the models presented in Figure 1. 
Panel a) presents the evolution with 
the scale factor of $\Omega_i$ for $\sin^2 2\theta_0=0.8$, $\Delta \xi_{\nu}=4$
($L_{\nu}=3.54$) and some values of $\Delta m^2$.
The same time evolution is presented in panel b) 
for 
$\Delta m^2=0.24$eV$^2$, $\Delta \xi_{\nu}=4$
($L_{\nu}=3.54$) and some values of $\sin^2 2\theta_0$.
In these cases, as  consequence of the conservation of the Hubble expansion rate 
for all values of the scale factor,  the time evolution of $\Omega_i$
do not  changes with the variation of $\sin^2 2\theta_0$.
Panel c) presents the evolution with 
the scale factor of $\Omega_i$ for $\Delta m^2=0.24$, $\sin^2 2\theta_0=0.45$ 
and some values of $\Delta \xi_{\nu}$/$L_{\nu}$.
Panels a1), b1) and c1) show 
the separate contributions from the two types of massive neutrinos
for the same cases of the above panels.
For all values of the scale factor, $\sum_i \Omega_i=1$.

Figure 3 shows the time evolution (in  synchronous gauge)
of the perturbed energy density of the density field components.
The left panels of Figure 3 present the evolution with the scale factor
of the perturbed energy density of different density field components
in the presence of neutrino oscillations
($\delta_{cdm}$ for CDM, $\delta_{bar}$ for baryons, $\delta_{rad}$
for radiation in the form of massless neutrino and photons, $\delta_{\nu_{\mu}}$
for $\nu_{\mu}$  plus ${\bar \nu}_{\mu}
$ and $\delta_{\nu_{\tau}}$ for
$\nu_{\tau}$ plus ${\bar \nu}_{\tau}$), 
for the mode $k \approx 0.2$Mpc$^{-1}$.
The right panels of Figure 3 present, for the same mode 
$k$, the evolution with the scale factor of the quantity
$\epsilon=\sigma_{\nu_1}-\sigma_{\nu_2}$ 
(proportional with the variation of the massive neutrinos 
shear stress), where $\sigma_{\nu_1}$ and $\sigma_{\nu_2}$ are:
\begin{eqnarray}
\sigma_{\nu_1}
= \sum_i \frac {\delta \rho_i} {\rho_i+P_i}, \hspace{1cm}  
\sigma_{\nu_2}
= \sum_j \frac {\delta \rho_j} {\rho_j+P_j} \hspace{3cm}
(i=\nu_{\tau},{\bar \nu}_{\tau} \;\;j=\nu_{\mu},{\bar \nu}_{\mu})
\end{eqnarray} 
Here $\rho_{\nu}$ and $P_{\nu}$ are the neutrino/antineutrino energy 
density and pressure and $\delta \rho_{\nu}$ is the perturbation 
of neutrino/antineutrino energy density. 
For all cases  the total neutrino mass is $m_{\nu}=0.6$eV, the total 
neutrino degeneracy parameter is $\xi_{\nu}=5$ and the cosmological 
model is the $\Lambda$CHDM model presented in Figure 1.
In panels a1) and a2) of Figure 3 
$\sin^2 2\theta_0=0.8$ and $\Delta \xi_{\nu}=4$ are fixed,
and $\Delta m^2=0.24$eV$^2$ (continuous line)
and  $\Delta m^2=0.12$eV$^2$ (dashed line). 
In panels b1) and b2) of the same figure 
$\Delta m^2=0.12$eV$^2$ and $\Delta \xi_{\nu}=4$ are
 fixed and $\sin^2 2\theta_0=0.8$ (continuous line),
$\sin^2 2\theta_0=0.23$ (dashed line).
We observe that the growth of
$\delta_{\nu_i}$ is suppressed when $\Delta m^2$ and $\sin^2 2\theta_0$
values are increased, while  the variation of neutrino 
shear stress gets larger with the increasing of these parameters.
In panels c1) and c2) of Figure 3
$\Delta m^2=0$, $\sin^2 2\theta_0=0$ 
and $\Delta \xi_{\nu}=4$ (continuous line) and $\Delta m^2=0$, 
$\sin^2 2\theta_0=0$ and $\Delta \xi_{\nu}=0$ (dashed line). For this last 
case, $\delta \rho_{\nu_{\mu}} \approx \delta \rho_{\nu_{\tau}}$
and $| \epsilon | < 10^{-2}$.
\begin{figure*}
\picplace{10cm}
\caption{The evolution with the scale factor (in synchronous gauge)
of the density perturbations  (left panels) 
and of $\epsilon=\sigma_{\nu_1}-\sigma_{\nu_2}$ (see also the text)
for the mode $k \approx 0.2$Mpc$^{-1}$. 
 Panels a1) and a2): $\sin^2 2\theta_0=0.8$, $\Delta \xi_{\nu}=4$ and
$\Delta m^2=0.24$eV$^2$ (continuous line) and $\Delta m^2=0.12$eV$^2$
(dashed line);
 Panels b1) and b2): $\Delta m^2=0.12$eV$^2$, $\Delta \xi_{\nu}=4$
 and $\sin^2 2\theta_0=0.8$
(continuous line) and $\sin^2 2\theta_0=0.23$ (dashed line);
Panels c1) and c2): $\Delta m^2=0$, $\sin^2 2\theta_0=0$,$\Delta \xi_{\nu}
=4$ (continuous line) and $\Delta m^2=0$, $\sin^2 2\theta_0=0$,
$\Delta \xi_{\nu}=0$ (dashed line).
For all cases  the total neutrino mass is $m_{\nu}=0.6$eV, the total
neutrino degeneracy parameter is $\xi_{\nu}=5$ and the cosmological
model is the $\Lambda$CHDM model presented in Figure 1.}
\end{figure*}

\section{Detection of neutrino oscillations with CMB experiments}

Different methods can be used to quantify the performance of a
CMB experiment in measuring  a given set of cosmological parameters.
We estimate the errors on the oscillation parameters $\Delta m^2$ and
$\sin^2 2\theta_0$ and the lepton asymmetry $L_{\nu}$ 
by using the Fisher information matrix approximation, widely employed in the
literature (e.g. Efstathiou \& Bond 1999, Popa et al. 1999).\\
The Fisher information matrix elements $F_{ij}$ measure the width and the
shape of the likelihood function around its maximum
(Efstathiou \& Bond 1999).
The minimum error that can be obtained
on a parameter $s_i$ when we need to determine all parameters
jointly, is given by:
\begin{equation}
\delta s_i = \sqrt{F^{-1}_{ii}},
\end{equation}
depending on the experimental parameter data set and the target model
and the class of considered  cosmological models.\\
If only the temperature anisotropy power spectrum $C_{Tl}$ is used,
the Fisher information matrix reads as (Efstathiou \& Bond 1999):
\begin{equation}
F_{ij}=\sum_{l} \frac{ \delta {\hat C}_{Tl} }{ \delta s_{i}} \cdot
Cov^{-1}( C^{2}_{Tl} ) \cdot \frac{ \delta C_{Tl} }{ \delta s_{j}}.
\end{equation}
If  both anisotropy and polarization power spectra are used,
the Fisher information matrix is given by (Zaldarriaga \& Seljak 1997a;
Zaldarriaga 1997):
\begin{equation}
F_{ij}=\sum_{l} \sum_{X,Y} \frac{\partial C_{Xl}}{\partial s_{i}}
              Cov^{-1}({\hat C}_{Xl},{\hat C}_{Yl})
                            \frac{\partial C_{Yl}}{\partial s_{j}},
\end{equation}
where: $X$ and $Y$ stands for $T$, $E$, $C$ and $B$ power spectra and
$Cov^{-1}$ is the inverse of the covariance matrix.
For the purpose of this work we assume only scalar modes,
then the relevant covariance matrix elements in equations (21) $\div$ (23)
are:

\begin{eqnarray}
 Cov({\hat C}^{2}_{Tl})&=&\frac{2}{(2l+1)f_{sky}}
       (C_{Tl}+w_T^{-1}B_{Tl}^{-2})^{2}, \nonumber%
\\
 Cov({\hat C}^{2}_{El})&=&\frac{2}{(2l+1)f_{sky}}
                    (C_{El}+w_{P}^{-1}B_{Pl}^{-2})^{2}, \nonumber%
\\
 Cov({\hat C}^{2}_{Cl})&=&\frac{2}{(2l+1)f_{sky}}
           [C^{2}_{Cl}+(C_{Tl}+w_T^{-1}B_{Tl}^{-2})
                        (C_{El}+w_{P}^{-1}B_{Pl}^{-2})], \nonumber%
\\
 Cov({\hat C}_{Tl}{\hat C}_{El})&=&\frac{2}{(2l+1)f_{sky}}C^{2}_{Cl},\nonumber%
\\
Cov({\hat C}_{Tl}{\hat C}_{Cl} )&=&\frac{2}{(2l+1)f_{sky}}
                    C_{Cl}(C_{Tl}+w_T^{-1}B_{Tl}^{-2}), \nonumber%
\\
Cov({\hat C}_{El}{\hat C}_{Cl} )&=&\frac{2}{(2l+1)f_{sky}}
C_{Cl}(C_{El}+w_{P}^{-1}B_{Pl}^{-2}).
\end{eqnarray}

Here we set $w_T=\sum_{c}w^T_{c}$ for anisotropy and $w_P=\sum_{c}w^P_{c}$ for polarization 
(with the sum performed over detector channels),
$w^T_{c}=(\sigma^T_c \theta_{c,pix})^{-2}$ and  $w^P_{c}=(\sigma^P_c \theta_{c,pix})^{-2}$,
where $\sigma^T_c$ and  $\sigma^P_c$  are the relative noise per pixel
for anisotropy and polarization channels (Knox 1995);  
$B^{2}_{Tl}=\sum_{c}B_{cl}^{2}w^T_{c}/w_T$ and $B^{2}_{Pl}=\sum_{c}B_{cl}^{2}w^P_{c}/w_P$
account for the beam smearing, $ B_{cl}^{2}=e^{-l(l+1)/l_{s}^{2}}$
is the Gaussian beam profile,
$l_{s}=\sqrt{8 \ln 2}(\theta_{c})^{-1}_{fwhm}$ and
$f_{sky}$ is the fraction of the sky used in the analysis.

Assessing  {\sc Planck} performances under realistic
assumptions is a very delicate task: both instrumental
systematic effects (e.g. Delabrouille 1998, De Maagt et al. 1998,
Maino et al. 1999, Burigana et al. 1998 and references therein)
and astrophysical contamination (e.g. Toffolatti et al. 1998,1999,
De Zotti et al. 1999a,b and references therein) have to be carefully
understood, reduced by
optimizing the telescope (Mandolesi et al. 1998b) and the instrumental
design (Bersanelli \& Mandolesi 1998, Lamarre et al. 1998)
and accurately subtracted in the data analysis.
For simplicity, we consider only the MAP and {\sc Planck}
``cosmological windows'', i.e. those frequencies that are
minimally affected
by foregrounds contaminations, and neglect the degradation
in CMB power spectrum recovering
introduced by foreground contaminations and instrumental effects.
As shown in Figure 4, in the cosmological channels extragalactic
source fluctuations are comparable to CMB ones only at $l > 10^3$,
both for temperature (left panel) and polarization (right panel)
anisotropies, whereas galactic fluctuations, which power spectrum 
overwhelms the extragalactic foreground one at $l \lsim
10^2$, are below the CMB ones at least far from the 
galactic plane.

Then, the possible imprinting of neutrino oscillations in the CMB 
anisotropy are not significantly masked from astrophysical contaminations
at frequencies $\sim 70 \div 200$ GHz. We exploit then separately
different sets of {\sc Planck} data:
the ``cosmological'' LFI channels alone (70 and 100 GHz channels),
the ``cosmological'' HFI channels alone (143 and 217 GHz channels
and 100 GHz but for the temperature fluctuations only)
and  both together.
For comparison, we consider also the set of data
from the two MAP ``cosmological'' channels at 60 and 90 GHz.
Table 1 lists the experimental parameters of the various experiments
that we considered in our calculations.
\begin{figure*}
\picplace{10cm}
\caption{Temperature (left panel) and polarization (right panel) power
spectra for the neutrino oscillation target model considered here (see
also the text) compared to astrophysical component power spectra:
the galactic temperature fluctuations at moderate galactic latitudes, sum
of synchrotron, free-free and dust fluctuations, and the extragalactic
source temperature fluctuations modelled as in Toffolatti et al. (1998) 
and De Zotti et al. (1999a) and the polarization fluctuations from
galactic synchrotron and dust emission and separately from
extragalactic radiosources and infrared sources modelled according to
De Zotti et al. (1999b) and references therein.
We report also the white noise nominal power spectrum 
for LFI and MAP, taking into account the effect of resolution loss 
at high $l$ due to the beam convolution.}
\end{figure*}

The {\sc Planck} and MAP data will be mainly used for determining the
cosmological parameters; we exploit then the Fisher matrix method by
evaluating together the uncertainties ($1\sigma$ errors)
of a simple set of cosmological parameters
and of the parameters characterizing the neutrino oscillation
model discussed here.  As well known, the errors quoted for the considered 
parameters depend on the experiment sensitivity but also 
in part on the assumed target model and 
the set of considered parameters. \\
We assume as target model a spatially flat $\Lambda$CHDM model
without oscillations
having: $\Omega_b=0.023$, $\Omega_{\Lambda}=0.7$,
$\Omega_c=1-\Omega_b-\Omega_{\nu}-\Omega_{\Lambda}$, $h_0=0.65$, $n_s=1$,
$m_{\nu}=m_{\nu_{\mu}}+m_{\nu_{\tau}}=0.6$eV, $\xi_{\nu}=\xi_{\nu_{\mu}}+\xi_{\nu_{\tau}}=5$,
$\Delta \xi_{\nu}=1$ ($L_{\nu}=2.13$),
$\Delta m^2=1.24 \times 10^{-2}$eV$^2$, and $\sin^2 2\theta_0=0$.
We also assume a scale invariant power spectrum, the presence only 
of the scalar modes having a spectral index $n_s=1$, 
and ignore the contribution
of the tensorial modes and reionisation effects.  
The tensorial modes have a small contribution only to the polarization 
signal at low order multipoles (Crittenden \& Turok 1995, Seljak 1997, 
Kamionkowski \& Kosowsky 1998) that will be difficult to detect against 
the polarized foregrounds. Also, the reionisation effects can help in
breaking the degeneracy only in the presence of the polarization signal
for a large value of the optical depth to Thomson scattering  
(Zaldarriaga, Spergel \& Seljak 1997). \\
We numerically compute the partial derivatives of the power spectra
with respect to each relevant  parameter $s_i$
by approximating them with the (possibly central) finite differences
between the two power spectra
obtained by varying with a quantity $ \pm \Delta s_i / 2$  the parameter
$s_i$ in consideration around its value $s_{i,0}$ assumed in the
considered target model. Typically, $\Delta s_i$ is taken in the range
between few percent, in order to have a step small enough
for an accurate evaluation of the derivative and large enough
for having quite sensitive and numerically stable variations of the power
spectrum.\\
Table 2 presents $1$-$ \sigma$ errors on the estimates of the
relevant cosmological parameters and neutrino oscillation
parameters obtained from the anisotropy alone
and anisotropy plus  polarization, for the experimental parameters
listed in Table 1 and  few values of $f_{sky}$.
The errors presented in Table 2 arise because of the geometrical
degeneracy (see Efstathiou \& Bond 1999 and the references therein)
that imposes limits on the determination of $h_0$ and $\Omega_{\Lambda}$,   
and the correlations existing among the energy density parameters and
 $n_s$ and $\sin^2 2\theta_0$.\\
These errors can be understood in terms 
of the ability of each experiment to determine the properties 
of Doppler peaks (Efstathiou \& Bond 1999): the location
of the first Doppler peak and the positions, heights and relative amplitudes of the 
first and subsidiary Doppler peaks. 
The Doppler peak positions $l_d$
 [$l_d \approx m \pi d_A(z)/r_s$,
where $d_A(z)$ is the  angular diameter distance  to
the last scattering, $r_s$ is the sound horizon distance and $m$ is 
the Doppler peak number] suffer the degeneracy between $\Omega_{\Lambda}$
and $h_0$ when $\Omega_m$ is fixed 
($\Omega_m=\Omega_b+\Omega_c+\Omega_{\nu}=0.3$). From  equation (19)
$\Omega_{\nu}$ value is given 
by  the $\Delta m^2$ and $L_{\nu}$ values when the total 
neutrino mass is fixed, and therefore $\Delta m^2$ and $L_{\nu}$ 
are also correlated.
The Doppler peak heights suffer the degeneracy between 
$\Omega_c$, $\Omega_b$, $\Omega_{\nu}$, $n_s$ and $\sin^2 2\theta_0$.\\
Clearly, the {\sc Planck} LFI, probing the subsidiary 
Doppler peaks structure to high multipoles $l \sim 1500$,  
will produce a significant improvement
with respect to MAP (sensitive up to $l \sim 1000$) in the determination
of all parameters reducing their uncertainties.
A further improvement can be achieved by exploiting the better sensitivity
and angular resolution (up to $l \sim 2000$) of {\sc Planck} HFI,
the full {\sc Planck} performance at very high $l$ being limitated essentially 
by the foreground contamination.
\begin{table*}[]
\caption[]{Summary of instrumental performances of future CMB
space missions in their ``cosmological channels''. 
For MAP and {\sc Planck} LFI we compute $\sigma^P_c=\sqrt{2} \sigma^T_c$ 
(Zaldarriaga \& Seljak 1997).}
\label{tableDummy}
\begin{flushleft}
\begin{tabular}{ccccccc}
\hline\\
Instrument  & $\nu$(GHz) & $\theta_{fwhm}$  & $\sigma^T_c/(\mu K/K)$&
$(w^T_c)^{-1}/10^{-15}$& $\sigma^P_c/(\mu K/K)$&$(w^P_c)^{-1}/10^{-15}$  \\
\hline
\ MAP                     & 60     & 21$'$   &  12.1 & 5.4&17.11&10.92  \\
\ (Bennet et al. 1996b)    & 90     & 12.6$'$ &  25.5 & 8.7&36.06&17.46  \\
\hline
\ {\sc Planck} LFI          & 70   & 14$'$   & 3.6  & 0.215&5.09&0.429 \\
\ (Mandolesi et al. 1998)   & 100  & 10$'$   & 4.3  & 0.156&6.08&0.312 \\
\hline
\ {\sc Planck} HFI          & 100   & 10.7$'$   & 1.7  & 0.028& & \\
\ (Puget et al. 1998)       & 143   & 8$'$      & 2.0  & 0.022&3.7&0.074 \\
                            & 217   & 5.5$'$    & 4.3  & 0.047&8.9&0.020 \\
\hline
\end{tabular}
\end{flushleft}
\end{table*}  
\begin{table*}[]
\caption[]{$1\sigma$ errors on the estimates of the cosmological parameters
and neutrino oscillation parameters obtained from the
power spectrum statistics.}
\label{tableDummy}
\begin{flushleft}
\begin{tabular}{ccccc|ccc}
\hline
 & &  \multicolumn{3}{c|}{Anisotropy}& \multicolumn{3}{c}{Anisotropy \&
polarization} \\
$ f_{sky}    $&             & 1.          & 0.8        & 0.65         & 1.
& 0.8   & 0.65 \\ \hline
          & $\delta n_{s} \times10^2$&1.44&1.61&1.79&1.30&1.46&1.62\\
          & $\delta \Omega_{\Lambda}/ \Omega_{\Lambda} \times 10^2$&6.01&6.
71&7.45&5.13&5.73&6.36\\
          & $\delta \Omega_{b}/ \Omega_b$&0.20&0.22&0.25&0.16&0.18&0.20\\
MAP       & $\delta h_{0}/ h_{0}$        &0.11&0.13&0.14&0.10&0.12&0.13\\
          & $\delta L_{\nu}/L_{\nu}$    &0.24&0.27&0.29&0.21&0.24&0.26\\
          & $\delta \Delta m^2 / \Delta m^2$&11.91&13.31&14.77&10.03&11.
22&12.45\\
          & $\delta \sin^2 2\theta_0$ &1.42&1.59&1.76&1.42&1.59&1.76\\      
\hline
          & $\delta n_{s} \times10^3$&4.69&5.24&5.81&2.46&2.76&3.06\\
 & $\delta \Omega_{\Lambda}/ \Omega_{\Lambda} \times 10^2$&2.24&2.50&2.78&0.99&1.11&1.23\\
          & $\delta \Omega_{b}/ \Omega_b\times 10^2 $&4.99&5.58&6.19&1.95&2
.18&2.42\\
{\sc Planck}-LFI
      &$\delta h_{0}/ h_{0} \times10^2$&3.26&3.65&4.05&1.66&1.86&2.06\\
      & $\delta L_{\nu}/L_{\nu}\times 10^2$&8.13&9.08&10.08&3.97&4.44&4.93\\
          &$\delta \Delta m^2 / \Delta m^2$&3.52&3.94&4.37&2.16&2.42&2.68\\
          &$\delta \sin^2 2\theta_0$&0.64&0.72&0.79&0.63&0.70&0.78\\
\hline
          &$\delta n_{s} \times10^3$&1.84&2.06&2.28&0.95&1.06&1.18\\
 & $\delta \Omega_{\Lambda}/ \Omega_{\Lambda} \times 10^3$&9.18&10.26&11.38&4.09&4.57&5.07\\
 &$\delta \Omega_{b}/ \Omega_b\times 10^2 $&1.80&2.01&2.23&0.64&0.72&0.80\\
{\sc Planck}-HFI
          &$\delta h_{0}/ h_{0} \times10^2$&1.74&1.95&2.16&0.64&0.72&0.79\\
          &$\delta L_{\nu}/L_{\nu}\times 10^2$&4.09&4.57&5.07&1.63&1.84&2.04\\
          &$\delta \Delta m^2 / \Delta m^2$&2.11&2.35&2.61&1.39&1.56&1.73\\
          &$\delta \sin^2 2\theta_0$&0.41&0.46&0.51&0.37&0.41&0.46\\
\hline
\end{tabular}
\end{flushleft}
\end{table*}
\begin{table*}[]
\caption[]{$1\sigma$ errors on the estimates of the neutrino oscillation
parameters and  lepton asymmetry
obtained from anisotropy and polarization power spectrum statistics.}
\label{tableDummy}
\begin{flushleft}
\begin{tabular}{ccccc}
\hline
$ f_{sky}    $&             & 1.          & 0.8        & 0.65  \\ \hline
        & $\delta L_{\nu}/L_{\nu}\times 10^2$&1.21&1.35&1.50\\
MAP     & $\delta \Delta m^2 / \Delta m^2$&4.08&4.56&5.06\\
        & $\delta \sin^2 2\theta_0\ $&0.68&0.77&0.85\\
\hline
        & $\delta L_{\nu}/L_{\nu} \times 10^3$&2.13&2.38&2.64\\
{\sc Planck} & $\delta \Delta m^2 / \Delta m^2$&0.79&0.88&0.97\\
        & $\delta \sin^2 2\theta_0$&0.13&0.19&0.21\\
\hline
\end{tabular}
\end{flushleft}
\end{table*}

On the other hand, 
other kinds of astronomical observations provide information
on the cosmological parameters, complementary to those derived from
CMB anisotropies.
Galaxy cluster observations limit 
the shape parameter $\Gamma$ and the normalization 
of the galaxy power spectrum $\sigma_8$
and, combined with the galaxy peculiar velocity fields,
provide informations on the matter density in the universe 
$\Omega_m$ (Dekel 1994, Strauss \& Willick 1995).
Weak gravitational
lensing provides also direct estimates of $q_0$ (see e.g Seljak 1998).
The ratio between baryon and total mass in clusters 
$\Omega_b/ \Omega_m \approx$ ($ 0.01$ - $0.02$) $h_0^{-2}$ 
(see e.g. Dodelson et al. 1996 and the references therein), 
informations from the BBN [$\Omega_b=0.02h_0^{-2}$ 
(Burles \& Tytler 1998)] and the Helium Lyman-Alpha Forest 
(Wadsley et al. 1999) constrain $\Omega_m$ or
$h_0$ if $\Omega_m$ is known from other kinds of measurements.
The complementarity between Type 1a supernovae luminosity distances and CMB 
anisotropy provides a  powerful tool
in breaking the degeneracy between $\Omega_k$ and $\Omega_{\Lambda}$
(see e.g. Perlmutter et al. 1998, Efstathiou \& Bond 1999,
Efstathiou 1999  and the references therein).
The ages of the oldest globular clusters
set lower limits on the age of the universe 
(see e.g. Chaboyer et al. 1998) and then on 
$\Omega_k$ and $\Omega_{\Lambda}$ if $\Omega_m$ is well constrained.
For a spatially flat universe this contributes to break the geometrical 
degeneracy in $h_0$ - $\Omega_{\Lambda}$ plane.
For CDM models with known baryonic content, 
measurements of $\Gamma$ from large scale structure, 
combined with independent determinations of $\Omega_m$, 
provide estimates on $h_0$ (Efstathiou \& Bond 1999).
Constraints on the scalar spectral index $n_s$
can be inferred also from 
limits on the primordial black hole abundance 
(Green, Liddle, \& Riotto 1997) 
and on the CMB spectral distortions (Hu, Scott, \& Silk 1994).

Recent CMB anisotropy measurements from {\sc Boomerang} and MAXIMA-1
have been in fact used jointly to other astronomical observations
to determine cosmological parameters through the so-called analysis with prior
(Lange et al. 2000, Balbi et al. 2000).
Of course, adding informations from other kind of observations
helps both in circumventing the degeneracy problem and in improving
the accuracy in the determination of the cosmological parameters.

Such a kind of analysis is out of the aim of this work.
We just estimate here lower limits on the MAP and {\sc Planck} 
sensitivity in determining the neutrino oscillation parameters,
by exploiting the Fisher matrix method, assuming that all the cosmological
parameters are known and estimating only the joint errors on $\Delta m^2$,
$\sin^2 2 \theta_0$, and $L_{\nu}$ that can be obtained from CMB
anisotropy plus polarization data. 

Table 3 presents these errors ($1\sigma$ errors)
for {\sc Planck} (LFI and HFI together) and MAP experiments;
together with the results shown in Table~2, it provides a range
of {\sc Planck} and MAP sensitivity to neutrino oscillation parameters.
Of course, in this case we find a significant reduction 
of the errors on the neutrino oscillation parameters 
for both {\sc Planck} and MAP.
These values can be also considered as estimates of lower limits 
on the errors on the neutrino oscillation
parameters, as they could be in principle stetted by
{\sc Planck} and MAP experiments.

Figure 5 presents few confidence 
regions of the neutrino oscillation parameters
space that can be potentially detected by the {\sc Planck} surveyor
and few allowed regions of $\Delta m^2$ and $\sin^2 2 \theta_0$  
at the same confidence level (CL)
obtained by MACRO Collaboration (Ambrosio et al. 1998).\\
According to the standard $\chi^2$ method (see e.g Particle Data Group 1998)
an input model defined by
the parameters ($\Delta m^2$, $\sin^2 2\theta_0$) is more likely to have
occurred higher is the probability to obtain $\chi^2$ values
larger then a specific value $\chi^2_0$.
The cumulative probability that
defines a certain confidence region on the parameters is given by:
\begin{equation}
CL(\chi^2_0)=\int_{\chi^2_0}
^{\chi^2_{max}} L(\chi^2,\Delta m^2$,$\sin^2 2\theta_0) d\chi^2,
\end{equation}
where  
$\chi^2_0=\chi^2_{min}+\Delta \chi^2$, with $\Delta \chi^2=4.61$, $9.21$
at $90\%$ and $99\%$ CL respectively (for a $\chi^2$ distribution
with two degrees of freedom).
The likelihood function in the equation (25) is defined as 
(Efstathiou \& Bond 1999):
\begin{eqnarray}
-2 {\rm ln} \left( \frac{L}{L_{max}} \right)= \sum_{l<l_{max}} 
\frac {(C_l(s)-C_l(s0))^2} {Cov ({\hat C}^2_{Tl})}, \nonumber
\end{eqnarray}
where $C_l(s_0)$ is the power spectrum for the target model, 
$C_l(s)$ is the power spectrum for the input model having
the same cosmological parameters as the target model and  
$\Delta m^2$ and $\sin^2 2\theta_0$
are in the intervals ($0 - 0.36$)eV$^2$ and ($0 -1$) respectively. \\
Panel a) of Figure 5 presents the  confidence regions that could be 
obtained with the  {\sc Planck}
surveyor for different fractions
of sky.
The target model is the $\Lambda CHDM$ model used
in the previous section.
Panel b) of the same figure presents the confidence 
regions obtained with {\sc Planck} 
for $f_{sky}=0.65$ and the same target model
and for another target model.\\
In each panel we also present the allowed regions  
of $\Delta m^2$ and $\sin^2 2 \theta_0$ as
derived from the MACRO experiment.\\
We conclude that exists  a significant overlap between the region of the
oscillation parameter space that can be potentially detected by {\sc Planck}
surveyor and that implied by the atmospheric neutrino oscillations data.

\begin{figure*}
\picplace{10cm}
\caption{Confidence regions of the neutrino oscillation parameter
space that can be potentially detected by {\sc Planck} surveyor
by using CMB anisotropy measurements.
Panel a): the allowed confidence regions at $99\%$ CL (A) and at 
$90\%$ CL (B) obtained for $f_{sky}=0.65$ (continuous lines) and
$f_{sky}=0.8$ (dashed lines). The target model has 
$\Delta m^2=1.24 \times 10^{-2}$eV$^2$ and $\sin^2 2\theta_0=0$ (see also the text).
Panel b): the allowed confidence regions at $99\%$ CL (A) and at 
$90\%$ CL (B) for $f_{sky}=0.65$. The target model has: 
$\Delta m^2=1.24 \times 10^{-2}$eV$^2$ and $\sin^2 2\theta_0=0$  
(continuous lines) and 
$\Delta m^2 \approx 10^{-4}$eV$^2$ and $\sin^2 2\theta_0=0$ (dashed lines).
In each panel we also present (dash-dotted lines) the allowed regions  
of $\Delta m^2$ and $\sin^2 2 \theta_0$ at 99$\%$ CL (A1)  and 90$\%$ CL (B1)
measured by the MACRO experiment.}
\end{figure*}

\section{Conclusions}

The imprint of relic degenerated neutrino
oscillations on the Cosmic Microwave Background (CMB)
angular power spectra is evaluated
in  a $\Lambda$CHDM model consistent with
the structure formation theories, allowing in the same time a pattern of
neutrino masses consistent with the atmospheric neutrino oscillations
data.\\
Under the assumption that the  total neutrino mass is known,
as derived by the galaxy redshift surveys
(Hu et al. 1998), we show that
oscillations occurring between two active degenerated neutrino flavors
leave  detectable imprints on the CMB anisotropy and polarization
power spectra. \\
By using the CMB anisotropy measurements, we find that
the relic degenerated neutrino oscillations
can be detected by {\sc Planck} surveyor at $1 \sigma$ level  
if the present value of the lepton asymmetry is $L_{\nu} \geq 9.7\times 
10^{-2}$,
the difference of the neutrino squared masses is
$\Delta m^2 \geq 2.91 \times 10^{-2}$eV$^2$ and the vacuum mixing angle is
$\sin^2 2\theta_0 \geq 0.46$, when we  consider
other cosmological parameters that simultaneously affect the CMB
angular  power spectra and a sky coverage of $f_{sky}=0.8$.
A slight improvement can be obtained by using the CMB anisotropy
and polarization data.

Assuming that all the other cosmological parameters may be known
from other kind of astronomical observations,
we estimate lower limits on future CMB anisotropy and polarization 
experiment sensitivity in determining the neutrino oscillation parameters
alone: as shown in Table 3, the errors on neutrino oscillation parameters
decrease at least by a factor two with respect to the
case in which no prior on cosmological parameters are assumed.
These values provide estimates of lower limits 
on the errors on the neutrino oscillation
parameters, as they could be in principle stetted by
{\sc Planck} and MAP experiments.

Finally, we compare the confidence 
regions of the neutrino oscillation parameter
space that can be potentially detected by the {\sc Planck} surveyor with 
the few allowed regions of $\Delta m^2$ and $\sin^2 2 \theta_0$  
obtained by MACRO Collaboration (Ambrosio et al. 1998).
We find a significant overlap between the region of the
oscillation parameter space that can be potentially detected by {\sc Planck}
.

\begin{acknowledgements}
We acknowledge the use of the CMBFAST Boltzmann code
(version 2.4.1) developed by U. Seljak and M. Zaldarriaga.
We gratefully thank the long-standing, very
fruitful collaboration on {\sc Planck} performances and
on foregrounds with M. Bersanelli, L. Danese, G. De Zotti,
D. Maino and L. Toffolatti.
It is a pleasure to acknowledge K. Enqvist, G. Giacomelli, R. Lopez, M.
Maris and G. Venturi
for useful and constructive discussions on neutrino oscillations.
We wish to thank the referee for constructive comments.
Part of this work was  supported by  NATO-CNR Fellowship Programme.
\end{acknowledgements}

\noindent
{\bf References}
\rref{Ambrosio, M. et al. (MACRO Collab.), 1998,Phys. Lett. B 434, 451}
\rref{Athanassopoulos, C. et al., 1998, Phys. Rev. Lett. 81, 1744}
\rref{Bahcall, J.N., Krastev, P.I. \& Smirnov, A.Yu., 1998, Phys. Rev.
D58, 096016}
\rref{Balbi, A. et al., 2000, preprint astro-ph/0005124}
\rref{Bennet, C. et al., 1996a, ApJ 464, L1}
\rref{Bennet, C. et al., 1996b, Amer. Astro. Soc. Meet., 88.05}
\rref{Bersanelli, M. \& Mandolesi, N., 1998, Astro. Lett. Comm., in press}
\rref{Bunn, E.F. \& White, M., 1997, ApJ. 480, 6}
\rref{Burigana, C. et al., 1998, A\&AS, 130, 551}
\rref{Burles, S. \& Tytler, D., 1998, Space Science Reviews, 84, 65}
\rref{Chaboyer, B., 1998, preprint astro-ph/9808200}
\rref{Crittenden, J.R. \& Turok, N., 1995, Phys. Rev. Lett. , 75, 2642} 
\rref{Dekel, A., 1994, Ann. Rev. Astron. Astrophys. 32, 319} 
\rref{Delabrouille, J., 1998, A\&AS, 127, 555}
\rref{De~Maagt, P., Polegre, A.M. \& Crone, G., 1998, 
{\sc Planck} -- Straylight
Evaluation of the Carrier Configuration, Technical Report ESA, 
PT-TN-05967, 1/0}
\rref{De~Zotti, G. et al., 1999a, 
in proc. {\it 3K Cosmology: EC-TMR Conference}; 
1999, L. Maiani, F. Melchiorri, and N. Vittorio eds., 
AIP Conf. Proc., pp. 204--223, astro-ph/9902103}
\rref{De~Zotti, G. et al., 1999b, New Astronomy, 4, 481}
\rref{Dodelson, S., Gyuk, G. \& Turner, M., 1994, Phys. Rev. D 49, 5068}
\rref{Dodelson, s., Gates, I.G. \& Turner, M., 1996, Science 274, 69}
\rref{Enqvist, K., Kainulainen, K. \& Thomson, M., 1992, Nucl. Phys. B 373,8}
\rref{Efstathiou, G. et al. 1999, MNRAS, 303, L47-52}
\rref{Efstathiou, G. \& Bond, J.R., 1999, MNRAS 304, 75}
\rref{Foot, R., Thomson, M.J. \& Volkas, R.R., 1996, Phys. Rev. D 53, 5349}
\rref{Foot, R. \& Volkas, R.R., 1997, Phys. Rev. D 56, 6653}
\rref{Foot, R. \& Volkas, R.R., 2000, Phys. Rev. D 61, 043507}
\rref{Freese, K. et al., 1983, Phys. Rev. D 27, 1689}
\rref{Fukuda, Y. et al. (Super-Kamiokande Collab.), 
1998, Phys. Rev. Lett. 81,1562}
\rref{Fukugita, M., Liu, G.C. \& Sugiyama, N. 1999, Phys. Rev. Lett. 84,
1082}
\rref{Gawiser, E. \& Silk, J., 1998, Science 280, 1405}
\rref{Gawiser, E. 2000, preprint astro-ph/0005475} 
\rref{G\'orski, K.~M. et al., 1994, ApJ 430, L89}
\rref{Green A.M., Liddle, A.R. \& Riotto, A., 1997, Phys. Rev. D56, 7559}
\rref{Hancock, S. et al. 1998, MNRAS, 294, L1}
\rref{Hannestad, S., 2000, preprint astro-ph/0005018}
\rref{Hu, W., Scott, D. \& Silk, J., 1994, ApJ, 430, 5}
\rref{Hu, W., Eisenstein, D.J. \& Tegmark M., 1998, Phys. Rev. Lett. 80,
5255}
\rref{Kamionkowski, M. \& Kosowsky A., 1998, Phys. Rev. D57, 685}
\rref{Kang, H. \& Steigman, G., 1992, Nucl. Phys. B 372, 494}
\rref{Kinney, W.H. \& Riotto, A., 1999, Phys. Rev. Lett. 83, 3366}
\rref{Knox, L., 1995, Phys. Rev. D52, 4307}
\rref{Lamarre, J.M. et al., 1998, Astro. Lett. Comm., in press}
\rref{Lange, A.E. et al., 2000, preprint astro-ph/0005004}
\rref{Larsen, G.B. \& Madsen, J., 1995, Phys. Rev. D 52, 4282}
\rref{Lesgourgues, J. \& Pastor, S., 1999, Phys. Rev. D 60, 103521}
\rref{Lesgourgues, J., Pastor, S. \& Prunet, S.  1999,
preprint hep-ph/9912363}
\rref{Ma, C. \& Bertschinger, E., 1995, Astrophys. J. 518, 2}
\rref{Maino, D. et al., 1999, A\&AS, 140, 1}
\rref{Mandolesi, N. et al., 1998a, {\sc Planck} LFI, A Proposal 
Submitted to the ESA}
\rref{Mandolesi, N. et al., 1998b, Astro. Lett. Comm., 
in press, astro-ph/9904135}
\rref{Mather, J.C., Fixsen, D.J., Shafer, R.A., 
Mosier, C., Wilkinson, D.T., ApJ, 512, 511}
\rref{Pal, P.B. \& Kar, K., 1999, Phys. Lett. B 451, 136}
\rref{Particle Data Group, 1998, The Europ. Phys. J. C 3, 1}
\rref{Perlmutter, S. et al. 1998, Nature, 391, 51}
\rref{Popa, L.A., Stefanescu, P. \& Fabbri, R. 1999, New Astronomy 4, 59}
\rref{Primack, J.R. et al., 1995, Phys. Rev. Lett. 74, 2160}
\rref{Primack, J.R. 1998,  Science 280, 1398}
\rref{Primack, J.R. \& Gross, M., 1998, to appear in the
Proceedings of the Xth Rencontres de Blois, ``The Birth of Galaxies'',
28 June - 4 July 1998, astro-ph/9810204}
\rref{Puget, J.~L. et al. 1998, HFI for the {\sc Planck} Mission, A Proposal
Submitted to the ESA}
\rref{Raffelt, G.G., 1996, in: ``Stars as Laboratories for
Fundamental Physics", eds. The University of Chicago Press,
Chicago \& London}
\rref{Riess, A.G. et al. 1998, Astron. J., 116, 1009}
\rref{Scott, D. \& White, M., 1994, in the Proceedings of the CWRU CMB
Workshop ``2 Years after COBE", eds. L. Krauss L. \& P. Kerman}
\rref{Seljak, U., 1997a, ApJ 482, 6}
\rref{Seljak, U., 1998, ApJ 506, 64}
\rref{Seljak, U. \& Zaldarriaga, M., 1996, ApJ 469, 437}
\rref{Smoot, G. et al., 1992, ApJ. 396, L1}
\rref{Steigman, G. et al., 1977, Phys. Lett. B66, 202}
\rref{Strauss, M. \& Willick, J., 1995, Phys. Repts. 261, 271}
\rref{Tegmark, M. \& Zaldarriaga, M., 2000, preprint astro-ph/0002091}
\rref{Toffolatti, L. et al., 1998, MNRAS 297, 117}
\rref{Toffolatti, L. et al., 1999, in proc. Intern. Conf.
{\it Microwave Foregrounds}; 1999, A. de Oliveira-Costa \& M. Tegmark eds.,
ASP Conf. Ser. Vol. 181, pp. 153--162, astro-ph/9902343}
\rref{Wadsley, J.W. et al., 1999, preprint astro-ph/9911394}
\rref{White, M., Gelmini, G. \& Silk, J., 1995, Phys. Rev. D 51, 2669}
\rref{Wright, E. et al., 1994, ApJ. 420, 1}
\rref{Zaldarriaga, M. \& Seljak, U., 1997, Phys. Rev. D 55, 1830}
\rref{Zaldarriaga, M. 1997, Phys. Rev. D 44, 1822}
\rref{Zaldarriaga, M., Spergel, D.N. \& Seljak, U., 1997, ApJ, 488, 1} 

\end{document}